\documentclass[aps,prb,twocolumn,showpacs,groupedaddress,floatfix]{revtex4}

\usepackage{graphicx}
\begin{document}


\title{Optical transitions and nature of Stokes shift in \\
spherical CdS quantum dots}

\author{D. O. Demchenko and Lin-Wang Wang}
\affiliation{Lawrence Berkeley National Laboratory, Berkeley, California 94720}
\date{\today}

\begin{abstract}
We study the structure of the energy spectra 
along with the character of the states participating in optical transitions 
in colloidal
CdS quantum dots (QDs) using the {\sl ab initio} accuracy
charge patching method combined with the 
folded spectrum
calculations of electronic structure of thousand-atom nanostructures. 
In particular, attention is paid to the nature of the 
large resonant Stokes shift observed in CdS quantum dots. 
We find that the top of the valence band state is bright, 
in contrast with the results of numerous {\bf k$\cdot$p} calculations, and 
determine the limits of applicability of the {\bf k$\cdot$p}
approach. The calculated electron-hole exchange splitting  suggests the 
spin-forbidden valence state may explain the
nature of the ``dark exciton'' in CdS quantum dots.

\end{abstract}


\maketitle
Study of optical properties of semiconductor quantum dots 
is largely driven by potential applications afforded by 
their size-dependent bandgap and exciton spectra, such as 
solar cells, \cite{solar} lasers, \cite{laser} and fluorescent tags
in byotechnology applications. \cite{tags1,tags2} 
In addition, quantum dots are excellent testing grounds 
for the applicability of various theoretical models. 
One of the most common features of quantum dots 
is the photoluminescence redshift 
relative to absorption (also called Stokes shift). 
There are various causes for the Stokes shift, most common is due to non-resonant 
absorption in the existence of large size variation in the sample. 
However even when the non-resonant component has been eliminated, e.g, by the florescence line
narrowing techniques (FLN) where only the largest quantum dots are excited, 
there can still exist the Stokes shift. This resonant Stoke shift is usually caused by 
a dark exciton ground state, but the nature of this dark ground state
can vary. A common dark ground state is caused by the exchange interaction 
between the electron and the hole producing a spin triplet ground state which is
spin-forbidden for optical transition. \cite{CdSe1,CdSe4}
Another possibility consists of the electron and hole having
different spatial envelope function symmetries. 
Since it is relatively rare, this spatial symmetry induced dark 
exciton ground state has been actively pursued in quantum dots. 
It has been indicated in previous theoretical and 
experimental studies that the exciton ground state
of a CdS quantum dot is one such candidate. \cite{CdS1,CdS_kp1} 
In this work, we will use ab initio accuracy 
calculation to re-investigate this problem. 

The use of the {\bf k$\cdot$p}
theory to theoretically investigate the structure of energy levels and
exciton states of spherical quantum dots has been very popular. 
\cite{kp1,kp2,kp3,kp4,kp5,kp6}
It has been shown by various authors, including ourselves, that \cite{CdS1,CdS_kp1,kp6} 
based on the {\bf k$\cdot$p} calculations,
the CdS QDs exciton ground state is an optically passive
``dark exciton'' for sufficiently small QDs because the hole ground state 
has an $P$-like envelope function. 
While the same have been found for CdSe and CdTe quantum dots, \cite{kp6} 
the results depend sensitively on the {\bf k$\cdot$p} parameters used. 
In contrast, for CdS quantum dots almost all reasonable {\bf k$\cdot$p}
parameters, and even different {\bf k$\cdot$p} Hamiltonian models 
(including cubic and wurtzite crystal field splitting)
predict the same 
$P$-like hole ground state, thus indicating the spatial symmetry induced dark exciton. 
With increasing quantum dot size the hole 
ground state within the {\bf k$\cdot$p} framework eventually becomes 
a bright 1S$_{3/2}$ state.
This result has seemingly been confirmed by the large values of observed Stokes shift in
CdS quantum dots, \cite{CdS1,CdS_kp1,CdS_kp2,CdS_kp3} the calculated 1P$_{3/2}$-1S$_{3/2}$ 
splitting values were shown to fit the observed values of resonant Stokes 
shift. \cite{CdS1}

While the {\bf k$\cdot$p} model had first successfully predicted 
the S-like state as the VBM in CdSe QDs \cite{kp4}, 
using a different set of {\bf k$\cdot$p} parameters 
generated from the pseudopotential bulk band structure 
it predicts the P-like state as the VBM, in contrast to the direct 
pseudopotential results \cite{CdSe_pseudo}. This shows that the {\bf k$\cdot$p} and the 
pseudopotential methods can predict different S/P ordering for the same set 
of {\bf k$\cdot$p} parameters.
Unfortunately, the pseudopotential Hamiltonian 
for CdS quantum dot was previously unavailable, therefore previous theoretical 
studies on CdS quantum dot have been limited to {\bf k$\cdot$p}-type calculations 
or tight-binding calculations. \cite{kp2,CdS1,CdS2,CdS_kp1,CdS_kp2,CdS_kp3} 
Recently, we have developed an ab initio accuracy charge patching method. 
This method can be used to calculate nanosystems of any given semiconductor material. 
A previous quantum dot/quantum wire  calculation using this method 
for thirteen different semiconductor materials \cite{LW_JB_dots} have demonstrated 
an excellent agreement between the calculated and experimental 
optical band gaps. Here, we will apply this method to re-investigate the dark 
exciton problem in CdS quantum dots.

The detailed recipe for our charge patching method calculations is 
given in Ref. \onlinecite{LW_JB_dots}, here we only give a briefly outline. 
The CdS quantum dots are assumed to have wurtzite structure with 
lattice constants $a=4.12$ \AA~ and $c=6.73$ \AA. 
The quantum dot effective radii $R_{eff}$ is determined from  
estimating the number of atoms in a sphere of radius $R_{eff}$, assuming
equal density of atoms in a quantum dot and the bulk. 
We are considering quantum dots (CdS)$_{43}$, (CdS)$_{92}$, (CdS)$_{183}$,
(CdS)$_{437}$, (CdS)$_{874}$, (CdS)$_{4586}$ 
(the subscript is the total number of Cd and S atoms), 
which have effective radii 
6.32 \AA, 8.15 \AA, 10.24 \AA, 13.70 \AA, 17.26 \AA, and 30 \AA~ respectively. 
The calculations were performed with the plane-wave pseudopotential
method, using local density approximation (LDA) and norm-conserving 
pseudopotentials. The plane wave energy cut-off of 35 Ry was used in all 
calculations. A well known LDA shortcoming of underestimating 
the bandgaps and the electron effective mass was corrected 
by modifying the nonlocal pseudopotentials 
(after Ref. \onlinecite{LW_JB_dots}),
such that the electron effective mass in the bulk is in a very good agreement
with experiment, i.e. $m_e=0.213m_0$ calculated here, versus $m_e=0.210m_0$
measured, \cite{reference} and the bandgap is partially corrected 
(it is not possible to correct both simultaneously), i.e. $E_g=2.16$ 
eV calculated versus $E_g=2.58$ eV measured \cite{reference} 
(uncorrected LDA yields \cite{LW_JB_dots}
$m_e=0.127m_0$ and $E_g=1.315$ eV). 
Spin-orbit coupling is included
and adjusted to yield $\Delta_{SO}$=0.068 eV corresponding to the 
experimental value. \cite{reference}
In order to eliminate the quantum dot surface dangling bond states
and keep the system charge neutral
we passivate the surface with pseudo-hydrogen 
following Ref. \onlinecite{LW_thousand}. A surface atom of valency
$m$ is passivated with a pseudo-atom with $Z=(8-m)/4$, therefore 
Cd and S are passivated with H($Z=1.5$) and H($Z=0.5$), respectively. 

The charge patching method \cite{LW_patching} is used in order to obtain the 
self-consistent quality real space charge distribution in the quantum dot, without
having to perform direct LDA calculations, which for the sizes of 
the systems considered here are prohibitively expensive. 
Here, only small prototype systems are computed self-consistently
for different atoms and their local environments to generate the 
motif charge densities . These charge motifs
are then used to assemble the total charge density for the entire 
quantum dot. Knowing the charge density the corresponding LDA potential 
is generated and the Hamiltonian of a given quantum dot is
constructed. The band edge eigenstates of this single particle 
Hamiltonian are then 
solved using the folded spectrum method. \cite{folded}

\begin{figure}[t]
\includegraphics[width=3.0 in,height=2.5 in]{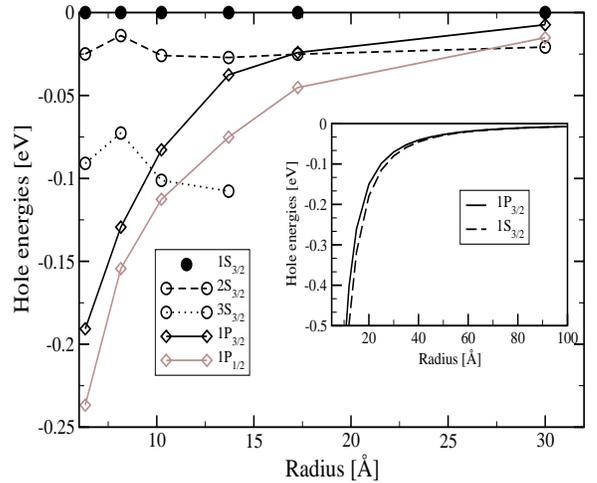}
\caption{Size dependence of the valence band energies 
without electron-hole Coulomb interaction
in CdS quantum dots, plotted relative to the 
top of the valence band, 1S$_{3/2}$ state. 
The inset shows the hole energies of the 1P$_{3/2}$ and  1S$_{3/2}$ states as a 
function of the quantum dot size, relative to the bulk
valence band maximum, calculated with the 
spherical {\bf k$\cdot$p} method. The effective mass parameters 
used are, $\gamma_1=2.31$, $\gamma_2=0.79$, and $\Delta_{SO}=0.068$ eV. 
\label{fig: E_noCoulomb}}
\end{figure}

Our calculated single particle eigenenergies for the valence band 
are shown in Fig.\ref{fig: E_noCoulomb}. 
Our calculation produces a 1S$_{3/2}$ state as the top
of the valence band, and the energies in Fig.\ref{fig: E_noCoulomb} 
are plotted relative to this state. 
For quantum dots in the region of  $R_{eff}>17$\AA~ the 1P$_{3/2}$ 
is the second hole state. For smaller quantum dots, there is another S$_{3/2}$ 
state above the 1P$_{3/2}$ state. 
Even for the largest QD calculated here, i.e. (CdS)$_{4586}$ with 
effective radius of $R_{eff}=30$\AA, we do not observe a 
S-to-P hole groundstate transition, and 1S$_{3/2}$ is still a groundstate. 
On the contrary, the $6\times6$ {\bf k$\cdot$p} theory in the 
spherical approximation, using the effective mass 
parameters derived from our ab initio bulk band structure 
($\gamma_1=2.31$, $\gamma_2=0.64$, $\gamma_3=0.89$, $\Delta_{SO}=0.068$ eV, 
$E_p=16.6$ eV, $E_g=2.16$ eV, $m_e=0.21m_0$, and in the spherical 
approximation $\gamma_{2,3}=0.79$), 
yields the 1P$_{3/2}$ groundstate, 
rather than the 1S$_{3/2}$ state, as shown in 
the inset to Fig. \ref{fig: E_noCoulomb}. 
For this set of parameters the P-to-S hole groundstate transition 
occurs at the QD effective radius of $\sim$100\AA. 
This is similar to other {\bf k$\cdot$p} calculations even 
though {\bf k$\cdot$p} parameters used might be somewhat 
different. \cite{kp2,kp6,CdS1} (Note, the caption of Fig. 5 in Ref. \onlinecite{CdS1}
is in error. The energy is given in the units of $\varepsilon_0=\gamma_1/(2R^2)$
not $\varepsilon_0=\gamma_1/(2R)^2$, and the $x$-axis is $0.529177 \times R$(nm), 
not $R$(nm).)
Note, that the $8\times8$ {\bf k$\cdot$p} formula should give qualitatively
similar results due to the large bandgap. 
Since the {\bf k$\cdot$p} predicts 1P$_{3/2}$ hole groundstate
for $R_{eff}<100$\AA, our calculations indicate that 
it is possible that the size range of 
$\sim$40\AA~$<R_{eff}<$ 100\AA, 
the 1P$_{3/2}$ is indeed the hole groundstate. 
However, the possible spatial symmetry induced dark exciton in that size range 
cannot be used to explain the experimentally observed Stokes shift, which is 
in the range of 15 - 70 meV, while the spatial dark exciton in that size range 
will have a Stokes shift of 0.6 - 6 meV, as predicted by the 
{\bf k$\cdot$p} model. 
The {\bf k$\cdot$p} method has been known to predict different ordering 
(and even omission) of the electronic levels in quantum dots \cite{CdSe_pseudo}
when compared to ab initio pseudopotential method. 
There has been a debate in the literature as to the proper comparison 
of the {\bf k$\cdot$p} and the ab initio methods \cite{debate}.
While describing the bulk 
reasonably well, the two methods can give qualitatively different results 
when applied to a nanocrystal 
(for a very detailed study see also Ref. \onlinecite{comparison}). 
Here we present another particular case of such disagreement.

To compare our results with experimental optical measurements, 
in addition to the single particle energies, 
we need to calculate the exciton energies. 
Exciton energies of optical transitions
in the strong confinement regime, where correlation effects are negligible,
can be calculated from the:
\begin{equation}
E_{ex}=\varepsilon_c - \varepsilon_v - E^{C}_{cv}-E^X_{cv},
\label{Exciton_energy}
\end{equation}
where, $\varepsilon_c$ and $\varepsilon_v$ are the single-particle 
conduction and valence states energies, respectively, and $E^{C}_{cv}$
is the electron-hole Coulomb energy, obtained as \cite{direct_coulomb}
\begin{equation}
E^{C}_{cv}=\int \int \frac{|\psi_c({\bf x}_1)|^2 |\psi_v({\bf x}_2)|^2}{\epsilon({\bf r}_1-{\bf r}_2)|{\bf r}_1-{\bf r}_2|} d {\bf x}_1 d{\bf x}_2
\label{Coulomb}
\end{equation}
where, ${\bf x}\equiv({\bf r},\sigma)$ includes both spatial ${\bf r}$ and 
spin $\sigma=\uparrow,\downarrow$ variables, $\epsilon({\bf r}_1-{\bf r}_2)$ 
is a position-dependent dielectric function (described below), and $\psi_c({\bf x})$ 
and $\psi_v({\bf x})$ are the wavefunctions for the conduction and 
valence states, respectively. 
The exciton energies can be further split by the electron-hole exchange 
interaction $E^X_{cv}$ in Eq.(\ref{Exciton_energy}).  
The exchange integral $E^X_{cv}$ is calculated as:
\begin{equation}
E^{X}_{cv}=\int \int \frac{\psi_{v}^*({\bf x}_1) \psi_{c}^*({\bf x}_2) \psi_{c}({\bf x}_1)  \psi_{v}({\bf x}_2)}{\epsilon({\bf r}_1-{\bf r}_2)|{\bf r}_1-{\bf r}_2|} d {\bf x}_1 d{\bf x}_2 .
\label{Exchange}
\end{equation}
In using the model dielectric function 
$\epsilon({\bf r}_1-{\bf r}_2)$ we follow the procedure outlined in 
Ref. \onlinecite{LW_screen}, where the short range exchange interaction
is essentially unscreened while the long range exchange interaction is 
screened significantly. 
Note that, to get the exchange splitting of the exciton energy, 
Eq.(\ref{Exciton_energy}) needs to be 
diagonalized among a few Kramers doublet spin configurations, 
which are degenerate
under the first three terms in Eq.(\ref{Exciton_energy}).

\begin{figure}[t]
\includegraphics[width=3.0 in,height=2.5 in]{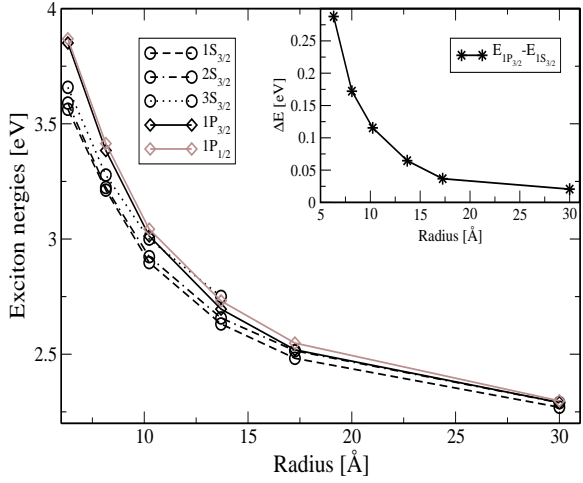}
\caption{Size dependence of the energies of excitons constructed from different 
valence band states
including electron-hole Coulomb interaction. 
The inset shows the energy difference between the 1P$_{3/2}$ and  1S$_{3/2}$ 
valence state constructed excitons
as a function of the CdS quantum dot size. 
\label{fig: E_Coulomb}}
\end{figure}

The calculated exciton energies using Eq.(\ref{Exciton_energy}) 
without the $E_{cv}^X$ term are shown in Fig.\ref{fig: E_Coulomb}.
The inset to Fig.\ref{fig: E_Coulomb} shows the 
1S$_{3/2}$ and 1P$_{3/2}$ energy difference which 
include electron-hole Coulomb interaction
as a function of the QD radius. 
The $E_{1P_{3/2}}-E_{1S_{3/2}}$ difference for the exciton has 
been enlarged compared to the single particle energies. 
This is because the 1S$_{3/2}$ hole state has larger
Coulomb interaction energy with the electron than the 1P$_{3/2}$ hole. 
Indeed, in the literature, there are reports of order changing in exciton 
states due to this Coulomb energy difference. \cite{CdS_kp2} In our 
case, it just makes the $P$-like hole state dark exciton less likely.

We then add in the exchange interaction, and calculate the exciton 
exchange splitting for our calculated 1S$_{3/2}$ hole ground state exciton, 
therefore producing a spin-forbidden dark exciton. 
While there is no debate about the screening in the electron-hole
Coulomb interaction, 
there exists a controversy regarding screening of the electron-hole exchange
interaction in excitons. 
For example, when calculating optical properties in Si quantum dots
\cite{unscreened_X} and hydrogenated Si clusters \cite{unscreened_X2} 
authors have calculated exciton energy structure with unscreened 
exchange interaction, following the argument that in the Bethe-Salpeter 
equation \cite{Sham_Rice} for the excitonic state the exchange term should be
unscreened, otherwise leading to improper diagrams. 
Recently, however, it was shown \cite{screen_Bethe} that when the 
two-particle Green's function is constructed from a small set of 
particle-hole states, it may be appropriate to screen the electron-hole 
exchange interaction. 
Therefore, in this work, along with the screened electron-hole exchange
splitting we also calculate unscreened exchange splitting for comparison.

\begin{figure}[t]
\includegraphics[width=3.0 in,height=2.5 in]{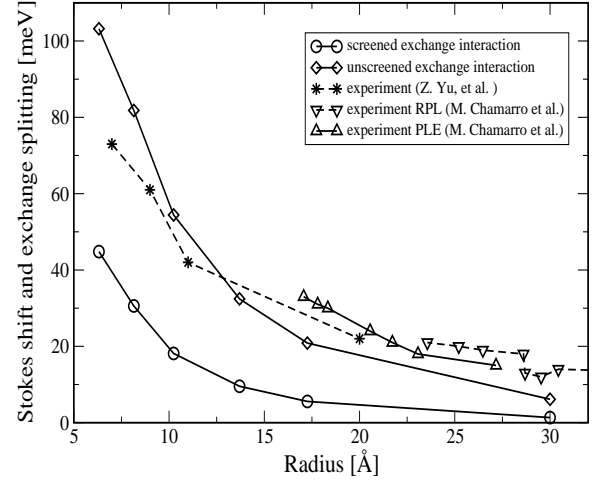}
\caption{Electron-hole exchange splitting in comparison with experimentally
measure resonant Stokes shift. Experimental data is deduced from 
Ref.\onlinecite{CdS1} (Z. Yu et al.) and from Ref.\onlinecite{CdS2}
(M. Chamarro et al.), measured with 
resonant photoluminescence (RPL) and photoluminescence excitation (PLE).
\label{fig: exchange}}
\end{figure}

The results are summarized in Fig. \ref{fig: exchange}, where
we plot the singlet-triplet exciton energy splitting for the screened 
exchange interaction, 
unscreened exchange interaction, and experimental data taken from 
Refs. \onlinecite{CdS1} and \onlinecite{CdS2}. 
The overall trend of the electron-hole exchange splitting agrees with 
the experimental data for the resonant Stokes shift. 
The experiment seems to favor the unscreened 
electron-hole exchange splitting results. 
This is in agreement with a previous tight-binding study \cite{CdS2} 
where the exchange 
interaction was also unscreened. On the other hand, 
there is a systematic discrepancy between the experiment and 
calculated values of the screened electron-hole exchange splitting. 
Overall, our results indicate that singlet-triplet splitting 
could be responsible for the observed values of the resonant Stokes 
shift in the experiment.  
In order to confirm this theoretical prediction we propose an experiment
in which the resonant Stokes shift is measured as a function of 
applied external magnetic field. 
The linear dependence of the Zeeman splitting on the external 
magnetic field could be detected from the Stokes shift measurement. \cite{CdSe1}

\begin{figure}[t]
\includegraphics[width=3.2 in,height=1.7 in]{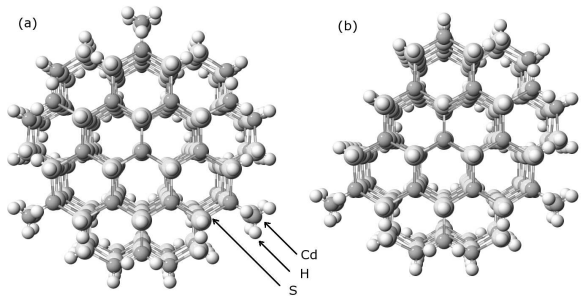}
\caption{Symmetric (a) and asymmetric (b) CdS quantum dots. 
The asymmetric dots exhibit a much (two orders of magnitude) smaller
radiative lifetimes for P-type hole states, compared with symmetric dots. 
\label{fig: sym_asym}}
\end{figure}

Last, to further study the possibility of the $P$-like hole 
induced spatial dark exciton, 
we have examined the geometry dependence of the spatial dark exciton 
radiative life time. 
Note that the quantum dot (wurtzite) structures considered so far 
are highly symmetric.  
Namely, there is a 120 degree rotation symmetry along the dot center $z$-axis 
(looking down in Fig. \ref{fig: sym_asym}).
However, quantum dots obtained in experiment will 
always have some structural imperfections, away from the perfect
symmetric shape. It is therefore interesting to study the influence of such
imperfection on the rate of optical transitions. 
Here we calculate radiative lifetimes $\tau$, following
Ref. \onlinecite{lifetime}.
Experimentally measured values for the radiative lifetimes
are \cite{CdS1} $\sim$180 ns for the slow 
component of the luminescence at the temperature of 10 K 
which corresponds to the radiative recombination of the 
optically passive state.  
In our calculations for the symmetric quantum dot [Fig.\ref{fig: sym_asym}(a)] we obtain
radiative lifetimes of 1 ns for $S$-like states, 
and very long ($> 1000$ ns) lifetime for P-like states, the latter is at least an order of 
magnitude larger than observed in experiment. 
We then calculated a similar sized quantum dot with its axial symmetry removed 
[Fig.\ref{fig: sym_asym}(b)]. As a result, the
radiative lifetime of the dark 1P$_{3/2}$ induced exciton
decreases sharply to $\sim$20 ns, much smaller than the observed experimental lifetime.  
Since it is hard to expect perfect symmetric structures in the
experiment, this result further supports the idea that the 
experimentally observed Stoke shift
is not caused by the spatial dark exciton, 
rather it might be caused by the exchange interaction 
induced spin-forbidden dark exciton. 
Note that the spin-forbidden dark exciton will not be eliminated by the
change of spatial shape of the quantum dot, and the CdS 
experimental lifetime of $\sim$180 ns (QD of $R=11$~\AA~ at $T=10$ K) is 
similar to the dark exciton lifetime in CdSe 
(210 ns for DQ of $R=13$~\AA~ at $T=12$ K) \cite{new_CdSe_lifetime} 
where it is known that the dark exciton is caused by the exchange splitting. 


In summary, 
we have performed calculations of exciton states, optical properties, and
electron-hole exchange splitting in CdS quantum dots. The results indicate
that the previous {\bf k$\cdot$p} method wrongly assigned the top of valence band  
for small ($R_{eff}< 30$\AA) CdS quantum dots. Our hole ground state is found 
to be bright S-state, rather than dark P-state predicted by the 
{\bf k$\cdot$p} method. As a consequence, the dark exciton is not spatially 
forbidden in CdS quantum dot. Our calculation indicates that the exchange splitting 
might be responsible  for the dark exciton and the observed Stoke shift. We recommend the 
magnetic field experiments to further resolve this issue.

We would like to thank Joshua Schrier for helping with the 
charge patching code, Byounghak Lee for useful discussions. 
This work was supported by the Director, Office of Energy Research, Office of Science, and
Division of Material Science, of the U.S. Department of Energy under Contract No.
DE-AC02-05CH11231. It used the resources of National 
Energy Research Scientific Computing Center (NERSC).

\end{document}